\documentstyle[12pt]{article}
\renewcommand{\baselinestretch}{1.2}

\begin{document}
\parskip=5pt plus 1pt minus 1pt
 
\begin{flushright}
{\bf hep-ph/9901329} \\
{\bf LMU-99-02} \\
{\bf WIS-99/01/Jan.DPP} \\
{\bf TAU-2547-99} \\
January 1999
\end{flushright}
 
\vspace{0.2cm}
 
\begin{center}
{\large\bf Flavor Symmetry, $K^0$-$\bar{K}^0$ Mixing and New Physics Effects\\
on $CP$ Violation in $D^{\pm}$ and $D^{\pm}_s$ Decays}
\end{center}
 
\vspace{0.5cm}
 
\begin{center}
{\bf Harry J. Lipkin} \footnote{Electronic address:
ftlipkin@wiswic.weizmann.ac.il} \\
{\sl Department of Particle Physics, Weizmann Institute
of Science, Rehovot 76100, Israel} \\
{\sl School of Physics and Astronomy,
Raymond and Beverly Sackler Faculty of Exact Sciences,
Tel Aviv University, Tel Aviv, Israel}
\end{center}
 
\begin{center}
{\bf Zhi-zhong Xing} \footnote{Electronic address:
xing@hep.physik.uni-muenchen.de} \\
{\sl Sektion Physik, Universit$\sl\ddot{a}$t
M$\sl\ddot{u}$nchen,
Theresienstrasse 37A, 80333 M$\sl\ddot{u}$nchen, Germany}
\end{center}
 
\vspace{2cm}
 
\begin{abstract}
Flavor symmetry and symmetry breaking, $K^0$-$\bar{K}^0$ mixing and possible
effects of new physics on $CP$ violation in weak decay modes
$D^{\pm}\rightarrow K_{\rm S,L}
+ X^{\pm}$, $(K_{\rm S,L} + \pi^0)^{~}_{K^*} + X^{\pm}$
(for $X =\pi, \rho, a_1$) and
$D^{\pm}_s \rightarrow K_{\rm S,L} + X^{\pm}_s$, $(K_{\rm S,L} +\pi^0)^{~}_{K^*}
+ X^{\pm}_s$ (for $X_s = K, K^*$) are analyzed.
Relations between $D^{\pm}$ and $D^{\pm}_s $ decay branching ratios    are
obtained from the $d \Leftrightarrow s$ subgroup of SU(3) and dominant
symmetry-breaking mechanisms are investigated. A $CP$ asymmetry of magnitude
$3.3\times 10^{-3}$ is shown to result in the standard model from
$K^0$-$\bar{K}^0$ mixing in the final-state. New physics affecting the doubly
Cabibbo-suppressed channels might either cancel this
asymmetry or enhance it up to the percent level.
A comparison between the $CP$ asymmetries in
$D^{\pm}_{(s)} \rightarrow K_{\rm S} X^{\pm}_{(s)}$ and
$D^{\pm}_{(s)} \rightarrow K_{\rm L} X^{\pm}_{(s)}$
can pin down effects of new physics.
\end{abstract}
 
 
\newpage
 
Effects of $CP$ violation in weak decays of $D$ mesons are
expected to be rather small, of order $10^{-3}$ or lower,
within the standard electroweak model \cite{Buccella}.
They can naturally be enhanced up to the $O(10^{-2})$ level,
if new physics beyond the standard model exists
in the charm-quark sector \cite{Bigi,Nir}.
Although no new physics model suggests direct $CP$ violation in charged
$D$-meson decays, a search is cheap and easy when decays of charge-conjugate
particles are measured \cite{Lipkin98}. 
Some efforts have so far been made to search for $CP$ violation in the $D$
system \cite{EX}. The experimental prospects are becoming brighter,
with the development of higher-luminosity
$e^+e^-$ colliders and hadron machines \cite{Kaplan,Liu}.
 
While a variety of mixing and $CP$-violating phenomena may
manifest themselves in neutral $D$-meson decays \cite{Xing97},
the charged
$D$-meson transitions provide a unique experimental opportunity
for the study of direct $CP$ violation \cite{Kaplan,Fry}. Some
phenomenological analyses of $CP$ asymmetries in
charged $D$-meson decay modes have been made
(see, e.g., Ref. \cite{Buccella,Lipkin98} and
Refs. \cite{Bigi95} -- \cite{Lipkin96}).
In particular, the importance and prospects of searching for
$CP$-violating new physics in the promising decays
$D^{\pm} \rightarrow K_{\rm S}
X^{\pm}$ and $D^{\pm} \rightarrow K_{\rm S} K_{\rm S} K^{\pm}$,
where $X^{\pm}$ denotes any charged hadronic state, have
been outlined in Ref. \cite{Lipkin98}.
 
In the present note we first consider flavor-symmetry relations between
corresponding $D^{\pm}$ and $D^{\pm}_s$ and then
discuss both the non-negligible $K^0$-$\bar{K}^0$ mixing effect
and the possible (significant) new
physics effect on $CP$ asymmetries in
$$
D^{\pm} \; \rightarrow \; K_{\rm S,L} ~ + ~ X^{\pm}
\eqno{\rm (1a)}
$$
(for $X=\pi , \rho , a_1$)
and
$$
D^{\pm}_s \; \rightarrow \; K_{\rm S,L} ~ + ~ X^{\pm}_s
\eqno{\rm (1b)}
$$
(for $X_s =K , K^*$)
decays. The similar decay modes involving the resonance $K^{*0}$
or $\bar{K}^{*0}$, i.e.,
$$
D^{\pm} \; \rightarrow \;  \left (K_{\rm S,L} + \pi^0 \right )_{K^*}
~ + ~ X^{\pm}
\eqno{\rm (2a)}
$$
and
$$
D^{\pm}_s \; \rightarrow \; \left (K_{S,L} + \pi^0 \right )_{K^*}
~ + ~ X^{\pm}_s
\eqno{\rm (2b)}
$$
are also considered.
Within the standard model
we show that $CP$ violation in these processes
comes mainly from $K^0$-$\bar{K}^0$ mixing and
may lead to a decay rate asymmetry of
magnitude $3.3\times 10^{-3}$.
Beyond the standard
model the $CP$ asymmetries are possible to reach the percent
level due to the enhancement from new physics.
A comparison between the $CP$ asymmetries in
$D^{\pm}_{(s)} \rightarrow K_{\rm S} X^{\pm}_{(s)}$
and $D^{\pm}_{(s)} \rightarrow K_{\rm L} X^{\pm}_{(s)}$
decays may pin down the involved new physics.
 
Within the standard electroweak model the transitions in Eqs. (1) and (2)
can occur through both the Cabibbo-allowed channels (Fig. 1)
and the doubly Cabibbo-suppressed channels (Fig. 2). The penguin
$c \rightarrow u$ transition cannot contribute to these decays.
The eight diagrams in Figs. 1 and 2 describe all the diagrams allowed by QCD
and the standard electroweak model if the quark lines are allowed to go
backward and forward in time, and arbitrary numbers of gluon exchanges are
included \cite{Close97}.
 
Each of the four diagrams in Fig. 1 is seen to go into one of the diagrams
of Fig. 2 under the $d \Leftrightarrow s$ transformation which interchanges
$d$ and $s$ flavors and is included in SU(3). Its use with the
flavor topology of Figs. 1 and 2 not only gives SU(3) symmetry relations
but also pinpoints the dominant sources of symmetry breaking. We first
note some relations between pairs of amplitudes
related by the $d \Leftrightarrow s$ transformation:
$$ 
\frac{|A(D^+\rightarrow K^{(*)0} X^+)|}
{|A(D^+_s\rightarrow \bar{K}^{(*)0} X^+_s)|}
\; = \; \frac{|V_{cd}V^*_{us}|}
{|V_{cs}V^*_{ud}|} \; = \; \frac{|A(D^+_s\rightarrow K^{(*)0} X^+_s)|}
{|A(D^+\rightarrow \bar{K}^{(*)0} X^+)|} \;\; ,
\eqno{\rm (3)}
$$
where $V_{ij}$ (for $i=u,c$ and $j=d,s$) are the Cabibbo-Kobayashi-Maskawa
(CKM) matrix elements.
The explicit assumptions needed to derive these relations are:
\begin{enumerate}
     \item $d$ and $s$ quarks couple equally to gluons at all stages of these
diagrams, including final state interactions.
     \item The properties of the weak interactions under the
$d \Leftrightarrow s$ transformation are given by the CKM matrix elements.
\end{enumerate}

That $d$ and $s$ quarks couple equally to gluons at short distances
in all the complicated quark diagrams seems reasonable. Thus further
$d \Leftrightarrow s$ symmetry breaking
occurs only at the hadron level via decay constants and form factors.
The $d \Leftrightarrow s$ transformation changes pions into kaons,
$\rho$'s into $K^*$'s, etc. The breaking produced by the corresponding
changes in masses, decay constants and form factors must be taken into account.
The form factors depend upon whether the $q \bar q$ pair is
pointlike or has a hadronic scale defined by the initial state wave function
or pair creation by gluons.
 
Branching ratio data for these decays may well become available with sufficient
precision to test SU(3) symmetry and symmetry breaking before they have
sufficient precision to show $CP$ violation or new physics. They can therefore
provide useful constraints on parameters and relative strengths of different
diagrams to help narrow the searches for crucial effects predicted by new
physics.
 
We now investigate the phases relevant to $CP$ violation.
For each decay mode under discussion,
its doubly Cabibbo-suppressed transition amplitude and
its Cabibbo-allowed one may have
different weak and strong phases.
We denote the ratio of two transition amplitudes for
$D^+\rightarrow K^{(*)0}X^+$ and $\bar{K}^{(*)0}X^+$
or $D^+_s\rightarrow K^{(*)0}X^+$ and $\bar{K}^{(*)0}X^+$
(before $K^0$-$\bar{K}^0$ mixing) as follows:
\setcounter{equation}{3}
\begin{eqnarray}
\frac{A(D^+\rightarrow K^{(*)0} X^+)}
{A(D^+\rightarrow \bar{K}^{(*)0} X^+)}
& = & R_d e^{{\rm i}\delta_d} \frac{V_{cd}V^*_{us}}
{V_{cs}V^*_{ud}} \; \; , \nonumber \\
\nonumber \\
\frac{A(D^+_s\rightarrow K^{(*)0} X^+_s)}
{A(D^+_s\rightarrow \bar{K}^{(*)0} X^+_s)}
& = & R_s e^{{\rm i}\delta_s} \frac{V_{cd}V^*_{us}}
{V_{cs}V^*_{ud}} \; \; ,
\end{eqnarray}
where $\delta_q$ and $R_q$ (for $q=d$ or $s$) denote
the strong phase difference and the ratio of real
hadronic matrix elements, respectively.
Under $d \Leftrightarrow s$ symmetry, $\delta_s = -\delta_d$
and $R_s = R^{-1}_d$.
 
The magnitudes of $R_d$ and $R_s$ can be estimated
with the help of the effective weak Hamiltonian
and the naive factorization approximation.
Neglecting the annihilation diagrams in Figs. 1 and 2, which are
expected to have significant form
factor suppression, we arrive at
\setcounter{equation}{4}
\begin{eqnarray}
\frac{1}{R_d} & = & 1 ~ + ~ \frac{a_1}{a_2} \cdot \frac{
\langle X^+|(\bar{u}d)^{~}_{V-A}|0\rangle \langle \bar{K}^{(*)0}|
(\bar{s}c)^{~}_{V-A}|D^+\rangle}
{\langle K^{(*)0}|(\bar{d}s)^{~}_{V-A}|0\rangle \langle X^+|
(\bar{u}c)^{~}_{V-A}|D^+\rangle} \; , \nonumber \\
\nonumber \\
R_s & = & 1 ~ + ~ \frac{a_1}{a_2} \cdot \frac{
\langle X^+_s|(\bar{u}s)^{~}_{V-A}|0\rangle \langle K^{(*)0}|
(\bar{d}c)^{~}_{V-A}|D^+_s\rangle}
{\langle \bar{K}^{(*)0}|(\bar{d}s)^{~}_{V-A}|0\rangle \langle X^+_s|
(\bar{u}c)^{~}_{V-A}|D^+_s\rangle} \; ,
\end{eqnarray}
where $a_1 \approx 1.1$ and
$a_2\approx -0.5$ are the effective Wilson coefficients
at the $O(m_c)$ scale \cite{Browder}.
We list the explicit results of $R_d$ and $R_s$ in Table 1,
in which the relevant
decay constants and form
factors are self-explanatory
and their values can be found from Refs. \cite{PDG98,BSW}.
 
These expressions show clearly that the relation  $R_s = R^{-1}_d$ is
violated only by the obvious $d \Leftrightarrow s$ breaking in the decay constants and
form factors, and holds in the limit where these decay constants and
form factors have the same values for all pairs related by
$d \Leftrightarrow s$; e.g.
$m_{\pi} = m^{~}_{K}$, $f_{\pi} = f_{K}$, $F^{DK}_0 (m^2_\pi)
= F^{DK}_0 (m^2_K)$,
$m_{\rho} = m^{~}_{K^*}$, $f_{\rho} = f_{K^*}$,
$F^{DK}_1 (m^2_\rho) = F^{D_sK}_1 (m^2_{K^*})$, etc.
 
The diagrams selected by factorization differ from other diagrams
only in the form factors. Thus relaxing the demand
for factorization can only introduce additional form factors and decay constants
into the expressions in Table 1 corresponding to the replacement of
color-favored couplings by color suppressed couplings.
 
The ballpark numbers obtained in Table 1 serve only
for illustration and show that the $d \Leftrightarrow s$ symmetry between
$R_s$ and $R^{-1}_d$ is reasonably acceptable for either
the two-pseudoscalar states or the pseudoscalar-vector (axialvector)
states. In general $|R_d| \sim O(1)$ and $|R_s| \sim O(1)$
are expected to be true, independent of the
dynamical details of these transitions.

We now consider the decays into final-states including $K_{\rm S}$ or 
$K_{\rm L}$, where the dominant $CP$ violation in the standard model  comes from 
the $K^0$-$\bar{K}^0$ mixing described by
\setcounter{equation}{5}
\begin{eqnarray}
|K_{\rm S}\rangle & = & p|K^{0}\rangle ~ + ~ q|\bar{K}^{0}\rangle
 \; , \nonumber \\
|K_{\rm L}\rangle & = & p|K^{0}\rangle ~ - ~ q|\bar{K}^{0}\rangle
 \; ,
\end{eqnarray}
where $p$ and $q$ are complex parameters.
To ensure the rephasing invariance of all
analytical results, we do not use  the
popular notation $q/p = (1-\epsilon)/
(1+\epsilon)$. As we shall see later on, the mixing-induced
$CP$ violation
\begin{equation}
\delta_K \; =\; \frac{|p|^2 ~ - ~ |q|^2}{|p|^2 ~ + ~ |q|^2}
\; \approx \; 3.3\times 10^{-3} \; ,
\end{equation}
which has been measured in semileptonic $K_{\rm L}$ decays
\cite{PDG98}, may play a significant role in the $CP$
asymmetries of $D^{\pm}$ and $D^{\pm}_s$ decays.
 
The ratios
of transition amplitudes for $D^-$ and $D^-_s$
decays can be read off from Eq. (4) with the
complex conjugation of relevant quark mixing matrix elements.
Then we obtain
\footnote{The formulas for the decays involving $K^{*0} \rightarrow
K^0 \pi^0 \rightarrow K_{\rm S} \pi^0$ and $\bar{K}^{*0} \rightarrow
\bar{K}^0 \pi^0 \rightarrow K_{\rm S} \pi^0$ are basically the same
as those for $D^{\pm} \rightarrow K_{\rm S}X^{\pm}$ or
$D^{\pm}_s \rightarrow K_{\rm S} X^{\pm}_s$ transitions, therefore
they will not be written down for simplicity.}
\begin{eqnarray}
\frac{A(D^+\rightarrow K_{\rm S} X^+)}{A(D^-\rightarrow
K_{\rm S} X^-)}
& = & \frac{(V_{cs}V_{ud}^*) q^* ~ + ~ R_d e^{{\rm i}\delta_d}
(V_{cd} V^*_{us}) p^*}
{(V^*_{cs} V_{ud}) p^* ~ + ~ R_d e^{{\rm i}\delta_d}
(V^*_{cd}V_{us}) q^*} \; \; , \nonumber \\
\nonumber \\
\frac{A(D^+_s\rightarrow K_{\rm S} X^+_s)}{A(D^-_s\rightarrow
K_{\rm S} X^-_s)}
& = & \frac{(V_{cs}V_{ud}^*) q^* ~ + ~ R_s e^{{\rm i}\delta_s}
(V_{cd} V^*_{us}) p^*}
{(V^*_{cs} V_{ud}) p^* ~ + ~ R_s e^{{\rm i}\delta_s}
(V^*_{cd}V_{us}) q^*} \; \; .
\end{eqnarray}
Although $q/p$ itself depends on the phase convention of
$|K^0\rangle$ and $|\bar{K}^0\rangle$ meson states, which
relies intrinsically on that of relevant quark fields \cite{Greenberg},
the following two quantities are
rephasing-invariant:
\begin{equation}
\frac{V_{cd}V^*_{us}}{V_{cs}V_{ud}^*} \cdot
\frac{p^*}{q^*} \; = \; r e^{+{\rm i}\phi} \; ,
~~~~~~~~
\frac{V^*_{cd}V_{us}}{V^*_{cs}V_{ud}} \cdot
\frac{q^*}{p^*} \; = \; \bar{r} e^{-{\rm i}\phi} \; .
\end{equation}
Note that $|q/p|$ deviates from
unity only at the $O(10^{-3})$ level, i.e., the order
of observable $CP$ violation in $K^0$-$\bar{K}^0$ mixing
\cite{PDG98}.
Therefore
$r = \bar{r} = \tan^2\theta_{\rm C} \approx 5\%$
is an excellent approximation, where
$\theta_{\rm C} \approx 13^{\circ}$ is the Cabibbo angle.
The $CP$ asymmetries of $D^{\pm}\rightarrow K_{\rm S} X^{\pm}$
and $D^{\pm}_s\rightarrow K_{\rm S} X^{\pm}_s$ transitions
can then be given as
\begin{eqnarray}
{\cal A}_d & = & \frac{|A(D^-\rightarrow K_{\rm S} X^-)|^2
- |A(D^+\rightarrow K_{\rm S} X^+)|^2}
{|A(D^-\rightarrow K_{\rm S} X^-)|^2
+ |A(D^+\rightarrow K_{\rm S} X^+)|^2}
\; \nonumber \\
\nonumber \\
& \approx &
\delta_K ~ + ~ 2 R_d \tan^2\theta_{\rm C} \sin\phi \sin\delta_d \; ,
\end{eqnarray}
and
\begin{eqnarray}
{\cal A}_s & = & \frac{|A(D^-_s\rightarrow K_{\rm S} X^-_s)|^2
- |A(D^+_s\rightarrow K_{\rm S} X^+_s)|^2}
{|A(D^-_s\rightarrow K_{\rm S} X^-_s)|^2
+ |A(D^+_s\rightarrow K_{\rm S} X^+_s)|^2}
\; \nonumber \\
\nonumber \\
& \approx &
\delta_K ~ + ~ 2 R_s \tan^2\theta_{\rm C} \sin\phi \sin\delta_s \; ,
\end{eqnarray}
where $\delta_K$ has been given in Eq. (7).
Clearly ${\cal A}_d$ or ${\cal A}_s$ consists of
two different contributions: that from $K^0$-$\bar{K}^0$
mixing in the final state, and that from the
interference between Cabibbo-allowed and doubly
Cabibbo-suppressed channels.
 
The smallness of $\delta_K$ implies that the
$K^0$-$\bar{K}^0$ mixing phase is nearly the
same as that in direct decays of $K^0$ and
$\bar{K}^0$ mesons \cite{Cohen}. Thus one may
take $q/p = (V_{us}V^*_{ud})/(V_{us}^*V_{ud})$
in the leading-order approximation. The
rephasing-invariant weak phase $\phi$ turns out to
be the largest outer angle of the unitarity triangle
$V_{cd}V^*_{ud} + V_{cs}V^*_{us} +
V_{cb}V^*_{ub} =0$ (see, e.g., Ref. \cite{Xing94})
and its magnitude can roughly
be constrained as follows:
\begin{equation}
\phi \; = \; \arg \left (\frac{V_{cd}V^*_{ud}}
{V_{cs}V^*_{us}} \right )
\; \geq \; \pi ~ - ~ \arctan \left |
\frac{V_{cb}V^*_{ub}}{V_{cd}V^*_{ud}} \right |
\; .
\end{equation}
In obtaining this bound, we have taken
$|V_{cd}V^*_{ud}| \approx |V_{cs}V^*_{us}|
\gg |V_{cb}V^*_{ub}|$ into account. We
find $\sin \phi \sim O(\leq 10^{-3})$ by
use of current experimental data on quark flavor
mixing \cite{PDG98}. This result, together with
$|R_q| \sim O(1)$ and $\tan^2\theta_{\rm C} \approx
5\%$, implies that the $CP$ asymmetry arising from
the interference between Cabibbo-allowed and
doubly Cabibbo-suppressed channels is
negligibly small in ${\cal A}_q$ (for $q=d$ or $s$).
One then concludes that
${\cal A}_s \approx {\cal A}_d \approx \delta_K$ holds to
a good degree of accuracy in the standard model.
Such $\delta_K$-induced
$CP$-violating effects can also be observed in
the semileptonic $D^{\pm}$ and $D^{\pm}_s$ decays
which involve $K_{\rm S}$ or $K_{\rm L}$ meson
via $K^0$-$\bar{K}^0$ mixing in the final
states \cite{Xing95}.
 
The conclusion drawn above may not be true if new
physics enters the decay modes under discussion.
As pointed out in Ref. \cite{Lipkin98}, one can
wonder about a kind of new physics which causes
the decay channels resembling the doubly Cabibbo-suppressed
ones in Fig. 2 but occurring through a new charged boson
or something more complicated than the $W$ boson.
The amplitudes of such new channels may remain to
be suppressed in magnitude (e.g., of the same order
as the doubly Cabibbo-suppressed amplitudes
in the standard model), but they are likely to
have significantly different weak and strong phases from the
dominant Cabibbo-allowed channels in Fig. 1 and
result in new $CP$-violating asymmetries via the
interference effects.
The strong phases are expected to be very different because of the presence
of meson resonances in the D mass region which affect the phases of the
doubly-suppressed $D^\pm$ and Cabibbo-allowed $D_s$ amplitudes 
while the Cabibbo-allowed
$D^\pm$ and doubly-suppressed $D_s$ decays feed exotic channels which have no
resonances \cite{Close97}.
Following this illustrative
picture of new physics, we modify the ratios of transition
amplitudes in Eq. (4) as follows:
\begin{eqnarray}
\frac{A(D^+\rightarrow K^0 X^+)}
{A(D^+\rightarrow \bar{K}^0 X^+)}
& = & R'_d e^{{\rm i}\delta'_d} \frac{U_{cd}U^*_{us}}
{V_{cs}V^*_{ud}} \; \; , \nonumber \\
\nonumber \\
\frac{A(D^+_s\rightarrow K^0 X^+_s)}
{A(D^+_s\rightarrow \bar{K}^0 X^+_s)}
& = & R'_s e^{{\rm i}\delta'_s} \frac{U_{cd}U^*_{us}}
{V_{cs}V^*_{ud}} \; \; ,
\end{eqnarray}
in which $\delta'_q$, $U_{cd}U^*_{us}$ and $R'_q$ (for $q=d$ or $s$)
stand for the effective
strong phase difference, the effective weak coupling and the
ratio of effective real
hadronic matrix elements, respectively.
Note that these quantities are composed of both the contribution
from doubly Cabibbo-suppressed channels in the standard model
and that from additional channels induced by new physics.
The same kind of new
physics might also affect $K^0$-$\bar{K}^0$ mixing, but
this effect can always be incorporated into the $p$ and $q$
parameters. In this case the rephasing-invariant
combinations in Eq. (9) become
\begin{equation}
\frac{U_{cd}U^*_{us}}{V_{cs}V_{ud}^*} \cdot
\frac{p^*}{q^*} \; = \; r' e^{+{\rm i}\phi'} \; ,
~~~~~~~~
\frac{U^*_{cd}U_{us}}{V^*_{cs}V_{ud}} \cdot
\frac{q^*}{p^*} \; = \; \bar{r}' e^{-{\rm i}\phi'} \; .
\end{equation}
Here $r' = \bar{r}'$ remains to be
a good approximation, but the magnitude of $r'$ (or $\bar{r}'$)
may deviate somehow from $\tan^2\theta_{\rm C} \approx 5\%$.
The $CP$ asymmetries of $D^{\pm}\rightarrow K_{\rm S} X^{\pm}$
and $D^{\pm}_s\rightarrow K_{\rm S} X^{\pm}_s$ decays, similar to
those obtained in Eqs. (10) and (11), read as
\begin{eqnarray}
{\cal A}'_d & \approx &
\delta_K ~ + ~ 2 R'_d r' \sin\phi' \sin\delta'_d \; ,
\nonumber \\
{\cal A}'_s & \approx &
\delta_K ~ + ~ 2 R'_s r' \sin\phi' \sin\delta'_s \; .
\end{eqnarray}
If $CP$ violation from the interference between the
channel induced by standard physics and that
arising from new physics
is comparable in magnitude with $\delta_K$
(i.e., at the $O(10^{-3})$ level) or dominant over
$\delta_K$ (i.e., at the $O(10^{-2})$ level), then
significant difference between ${\cal A}'_s$ and
${\cal A}'_d$ should in general appear. This follows
that $\delta'_s \neq \delta'_d$ and they may
even have the opposite signs.
For illustration, we plot the dependence
of ${\cal A}'_d$
and ${\cal A}'_s$ on $\phi'$ in Fig. 3 with the
typical inputs $r'=0.04$, $R'_d \sin\delta'_d =0.3$ and
$R'_s \sin\delta'_s =-0.5$.
Clearly it is worthwhile to measure both
asymmetries, and a comparison between them will
be helpful to examine SU(3) symmetry and
probe possible new physics effects.
 
The $CP$-violating asymmetries in $D^{\pm}\rightarrow K_{\rm L}X^{\pm}$
and $D^{\pm}_s\rightarrow K_{\rm L}X^{\pm}_s$ decays
can directly be obtained from
those in Eqs. (10), (11) and (15) through the replacements
$R_q \rightarrow -R_q$ and $R'_q \rightarrow -R'_q$
(for $q=d$ or $s$), as ensured by $CPT$ symmetry
in the total width \cite{Lipkin95} .
The point is simply that
$q/p$ becomes $-q/p$, if one moves from the
$K_{\rm S}X^{\pm}_{(s)}$
state to the $K_{\rm L}X^{\pm}_{(s)}$ state.
The difference between the $CP$ asymmetries in
$D^{\pm}_{(s)}\rightarrow K_{\rm S} X^{\pm}_{(s)}$
and $D^{\pm}_{(s)} \rightarrow K_{\rm L} X^{\pm}_{(s)}$
turns out to be
\begin{eqnarray}
{\cal A}'_d (K_{\rm S}) ~ - ~ {\cal A}'_d (K_{\rm L})
& \approx & 4 R'_d r' \sin\phi' \sin\delta'_d \; ,
\nonumber \\
{\cal A}'_s (K_{\rm S}) ~ - ~ {\cal A}'_s (K_{\rm L})
& \approx & 4 R'_s r' \sin\phi' \sin\delta'_s \; .
\end{eqnarray}
This implies an interesting possibility to pin down the involved
new physics, which significantly violates $CP$ invariance.
For either $D^{\pm}$ or
$D^{\pm}_s$ decays, we may explicitly conclude that
the $CP$-violating effect induced by new physics is
\begin{itemize}
\item	vanishing or
negligibly small, if the relationship ${\cal A}'_q (K_{\rm S})
\approx {\cal A}'_q (K_{\rm L})
\approx \delta_K$ is observed in experiments;
 
\item	comparable in magnitude with
the $\delta_K$-induced $CP$ violation, if
$|{\cal A}'_q (K_{\rm S})| \gg |{\cal A}'_q (K_{\rm L})|$
(or vice versa) is experimentally detected;
 
\item	dominant over the
$\delta_K$-induced $CP$ asymmetry, if
${\cal A}'_q (K_{\rm S}) \approx - {\cal A}'_q
(K_{\rm L})$ (of order $10^{-2}$) is measured in experiments.
\end{itemize}
These conclusions are quite general and they should
also be valid for other types of
new physics involved in the charm-quark sector.
 
In view of current experimental data \cite{PDG98},
the branching ratios of $D^{\pm} \rightarrow
K_{\rm S,L} + (\pi^{\pm}, \rho^{\pm}, a^{\pm}_1)$
are estimated to be about $1.4\%$, $3.3\%$ and $4.0\%$,
respectively. The branching ratios of $D^{\pm}_s
\rightarrow K_{\rm S,L} + (K^{\pm}, K^{*\pm})$
amount approximately to $1.8\%$ and $2.2\%$, respectively
\footnote{The decay modes involving the $\bar{K}^{*0}$ (or $K^{*0}$)
resonance may have smaller branching ratios, as both
$\bar{K}^{*0}\rightarrow \bar{K}^0\pi^0$ and $\bar{K}^{*0} \rightarrow
K^-\pi^+$ are allowed but only the former is relevant to our purpose.
For example, the branching ratio of $D^+\rightarrow (K_{\rm S,L} + \pi^0)
^{~}_{K^*} + \pi^+$ is expected to be $3.2\times 10^{-3}$ from
current data \cite{PDG98}, about 1/5 of the branching ratio of
$D^+\rightarrow K_{\rm S,L} + \pi^+$.}.
It is therefore possible to measure the $\delta_K$-induced
$CP$ asymmetries in these decay modes with about
$10^{7-8}$ $D^{\pm}$ or $D^{\pm}_s$ events. If
new physics enhances the asymmetries up to the
percent level, then clean signals of $CP$ violation
can be established with only about $10^6$ $D^{\pm}$
or $D^{\pm}_s$ events.
 
\vspace{0.5cm}
\underline{Acknowledgement:} ~
It is a pleasure to thank Jeffrey Appel, Edmond Berger, Karl Berkelman, Ikaros
Bigi, John Cumalat, Harald Fritzsch, Yuval Grossman, 
Yosef Nir, J. G. Smith and Yue-liang Wu for helpful discussions and comments.
This work was partially supported by the German-Israeli Foundation
for Scientific Research and Development (GIF).

\newpage
 
\begin{table}[t]
\caption{Estimation of $R_d$ and $R_s$ in the naive factorization
approximation. Note that
the expressions for $D^+ \rightarrow K^{*0} \rho^+$
vs $\bar{K}^{*0} \rho^+$ and $K^{*0} a^+_1$ vs $\bar{K}^{*0} a^+_1$
as well as $D^+_s \rightarrow K^{*0} K^{*+}$ vs $\bar{K}^{*0} K^{*+}$
decays are too complicated to be listed here.}
\vspace{-0.2cm}
\begin{center}
\begin{tabular}{lllll} \\ \hline\hline
$D^+$ decays     &~~& Expression of $R^{-1}_d$	&~~& Value of $|R_d|^{-1}$
\\ \hline \\
$K^0\pi^+$ vs $\bar{K}^0\pi^+$
&& $\displaystyle 1 + \frac{a_1}{a_2} \cdot \frac{f_{\pi}}{f_K}
\cdot \frac{F^{DK}_0(m^2_{\pi})}{F^{D\pi}_0(m^2_K)} \cdot
\frac{m^2_D - m^2_K}{m^2_D - m^2_\pi}$	
&& $0.79$ \\ \\
$K^0\rho^+$ vs $\bar{K}^0\rho^+$
&& $\displaystyle 1 + \frac{a_1}{a_2} \cdot \frac{f_{\rho}}{f_K}
\cdot \frac{F^{DK}_1(m^2_{\rho})}{A^{D\rho}_0(m^2_K)}$
&& $2.7$ \\ \\
$K^0a^+_1$ vs $\bar{K}^0a^+_1$
&& $\displaystyle 1 + \frac{a_1}{a_2} \cdot \frac{f_{a_1}}{f_K}
\cdot \frac{F^{DK}_1(m^2_{a_1})}{A^{Da_1}_0(m^2_K)}$
&& $3.9$ \\ \\
$K^{*0}\pi^+$ vs $\bar{K}^{*0}\pi^+$
&& $\displaystyle 1 + \frac{a_1}{a_2} \cdot \frac{f_\pi}{f_{K^*}}
\cdot \frac{A^{DK^*}_0(m^2_\pi)}{F^{D\pi}_1 (m^2_{K^*})}$
&& $0.54$ \\ \\
\hline\hline
$D^+_s$ decays     &~~& Expression of $R_s$	&~~& Value of $|R_s|$
\\ \hline \\
$K^0K^+$ vs $\bar{K}^0K^+$
&& $\displaystyle 1 + \frac{a_1}{a_2}$
&& $1.2$ \\ \\
$K^0K^{*+}$ vs $\bar{K}^0K^{*+}$
&& $\displaystyle 1 + \frac{a_1}{a_2} \cdot \frac{f_{K^*}}{f_K}
\cdot \frac{F^{D_sK}_1(m^2_{K^*})}{A^{D_sK^*}_0(m^2_K)}$
&& $2.5$ \\ \\
$K^{*0} K^+$ vs $\bar{K}^{*0} K^+$
&& $\displaystyle 1 + \frac{a_1}{a_2} \cdot \frac{f_K}{f_{K^*}}
\cdot \frac{A^{D_sK^*}_0(m^2_K)}{F^{D_sK}_1(m^2_{K^*})}$
&& $0.36$ \\ \\
\hline\hline
\end{tabular}
\end{center}
\end{table}
 
\newpage
 
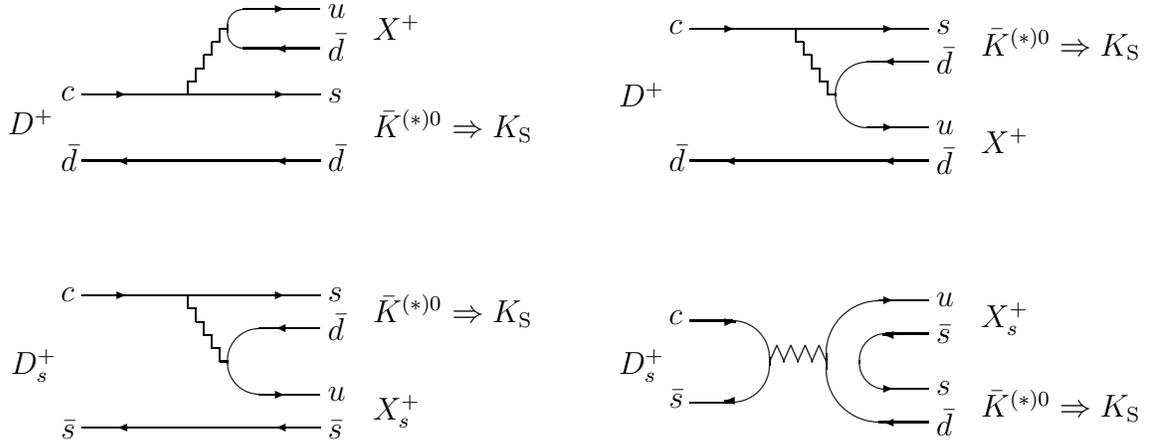
\begin{figure}
\begin{picture}(400,250)
\put(70,215){\line(1,0){90}}
\put(62,213){$c$}
\put(62,186){$\bar{d}$}
\put(42,200){$D^{+}$}
\put(163,212){$s$}
\put(163,186){$\bar{d}$}
\put(163,244){$u$}
\put(163,228){$\bar{d}$}
\put(180,236){$X^{+}$}
\put(180,199){$\bar{K}^{(*)0}\Rightarrow K_{\rm S}$}
\put(70,190){\line(1,0){90}}
\put(160,240){\oval(70,15)[l]}
\put(145,247.5){\vector(1,0){2}}
\put(145,232.5){\vector(-1,0){2}}
\put(85,215){\vector(1,0){2}}
\put(145,215){\vector(1,0){2}}
\put(85,190){\vector(-1,0){2}}
\put(145,190){\vector(-1,0){2}}
\multiput(110,215)(3,5){5}{\line(0,1){5}}
\multiput(107,215)(3,5){6}{\line(1,0){3}}
\put(300,240){\line(1,0){90}}
\put(300,190){\line(1,0){90}}
\put(292,238){$c$}
\put(292,186){$\bar{d}$}
\put(273,210.5){$D^{+}$}
\put(393,238){$s$}
\put(393,185){$\bar{d}$}
\put(393,223){$\bar{d}$}
\put(393,200){$u$}
\put(410,193){$X^{+}$}
\put(410,230){$\bar{K}^{(*)0}\Rightarrow K_{\rm S}$}
\put(390,215){\oval(70,25)[l]}
\put(315,240){\vector(1,0){2}}
\put(315,190){\vector(-1,0){2}}
\put(375,240){\vector(1,0){2}}
\put(375,190){\vector(-1,0){2}}
\put(375,227.5){\vector(-1,0){2}}
\put(375,202.5){\vector(1,0){2}}
\multiput(340,240)(3,-5){5}{\line(0,-1){5}}
\multiput(337,240)(3,-5){6}{\line(1,0){3}}
\end{picture}
 
\begin{picture}(400,100)
\put(70,240){\line(1,0){90}}
\put(70,190){\line(1,0){90}}
\put(62,238){$c$}
\put(62,186){$\bar{s}$}
\put(43,210.5){$D^{+}_s$}
\put(163,238){$s$}
\put(163,186){$\bar{s}$}
\put(163,223){$\bar{d}$}
\put(163,200){$u$}
\put(180,193){$X^{+}_s$}
\put(180,230){$\bar{K}^{(*)0}\Rightarrow K_{\rm S}$}
\put(160,215){\oval(70,25)[l]}
\put(85,240){\vector(1,0){2}}
\put(85,190){\vector(-1,0){2}}
\put(145,240){\vector(1,0){2}}
\put(145,190){\vector(-1,0){2}}
\put(145,227.5){\vector(-1,0){2}}
\put(145,202.5){\vector(1,0){2}}
\multiput(110,240)(3,-5){5}{\line(0,-1){5}}
\multiput(107,240)(3,-5){6}{\line(1,0){3}}
\put(292,229){$c$}
\put(292,197.5){$\bar{s}$}
\put(273,210.5){$D^{+}_s$}
\put(393,236){$u$}
\put(393,187){$\bar{d}$}
\put(393,202.5){$s$}
\put(393,222){$\bar{s}$}
\put(410,228){$X^{+}_s$}
\put(410,194){$\bar{K}^{(*)0}\Rightarrow K_{\rm S}$}
\put(300,215){\oval(60,31)[r]}
\put(390,215){\oval(52,21)[l]}
\put(390,215){\oval(77,46)[l]}
\put(315,230){\vector(1,0){2}}
\put(315,200){\vector(-1,0){2}}
\put(375,225){\vector(-1,0){2}}
\put(375,204.5){\vector(1,0){2}}
\put(375,238){\vector(1,0){2}}
\put(375,192){\vector(-1,0){2}}
\multiput(329.2,215)(5.3,0){4}{\small $\wedge$}
\end{picture}
\vspace{-6.2cm}
\caption{Cabibbo-allowed channels for
$D^{+}\rightarrow K_{\rm S} X^+$,
$(K_{\rm S} \pi^0)^{~}_{K^*} X^+$
and $D^+_s \rightarrow K_{\rm S} X^+_s$, $(K_{\rm S} \pi^0)^{~}_{K^*} X^+_s$
decays in the standard model.
Here $X = \pi$, $\rho$ or $a_1$,
$X_s = K$ or $K^*$, and $K_{\rm S}$ can
be replaced by $K_{\rm L}$.}
\end{figure}
 
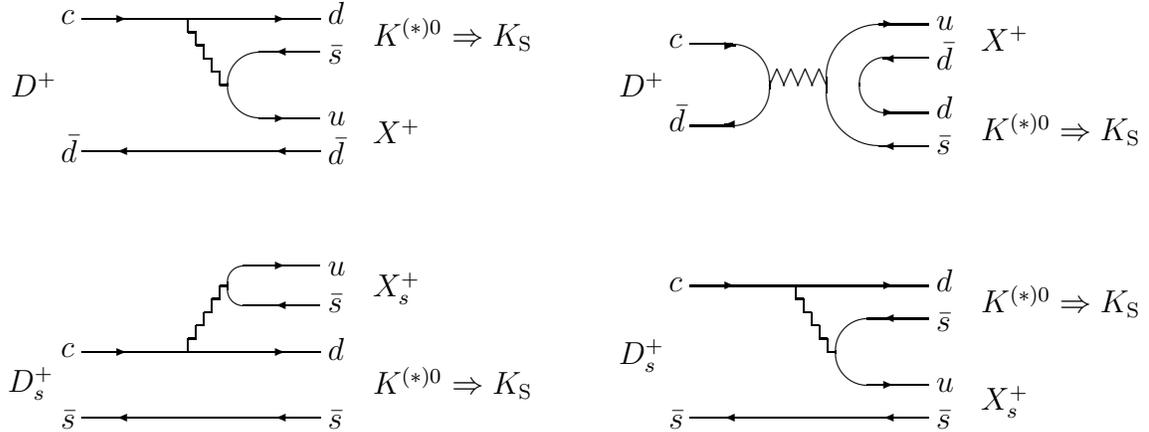
\begin{figure}
\begin{picture}(400,200)
\put(70,240){\line(1,0){90}}
\put(70,190){\line(1,0){90}}
\put(62,238){$c$}
\put(62,186){$\bar{d}$}
\put(43,210.5){$D^{+}$}
\put(163,238){$d$}
\put(163,185){$\bar{d}$}
\put(163,223){$\bar{s}$}
\put(163,200){$u$}
\put(180,193){$X^{+}$}
\put(180,230){$K^{(*)0}\Rightarrow K_{\rm S}$}
\put(160,215){\oval(70,25)[l]}
\put(85,240){\vector(1,0){2}}
\put(85,190){\vector(-1,0){2}}
\put(145,240){\vector(1,0){2}}
\put(145,190){\vector(-1,0){2}}
\put(145,227.5){\vector(-1,0){2}}
\put(145,202.5){\vector(1,0){2}}
\multiput(110,240)(3,-5){5}{\line(0,-1){5}}
\multiput(107,240)(3,-5){6}{\line(1,0){3}}
\put(292,229){$c$}
\put(292,197.5){$\bar{d}$}
\put(273,210.5){$D^{+}$}
\put(393,236){$u$}
\put(393,189){$\bar{s}$}
\put(393,202.5){$d$}
\put(393,221){$\bar{d}$}
\put(410,228){$X^{+}$}
\put(410,194){$K^{(*)0}\Rightarrow K_{\rm S}$}
\put(300,215){\oval(60,31)[r]}
\put(390,215){\oval(52,21)[l]}
\put(390,215){\oval(77,46)[l]}
\put(315,230){\vector(1,0){2}}
\put(315,200){\vector(-1,0){2}}
\put(375,225){\vector(-1,0){2}}
\put(375,204.5){\vector(1,0){2}}
\put(375,238){\vector(1,0){2}}
\put(375,192){\vector(-1,0){2}}
\multiput(329.2,215)(5.3,0){4}{\small $\wedge$}
\end{picture}
 
\begin{picture}(400,100)
\put(70,215){\line(1,0){90}}
\put(62,213){$c$}
\put(62,186){$\bar{s}$}
\put(42,200){$D^{+}_s$}
\put(163,212){$d$}
\put(163,186){$\bar{s}$}
\put(163,244){$u$}
\put(163,230){$\bar{s}$}
\put(180,236){$X^{+}_s$}
\put(180,199){$K^{(*)0}\Rightarrow K_{\rm S}$}
\put(70,190){\line(1,0){90}}
\put(160,240){\oval(70,15)[l]}
\put(145,247.5){\vector(1,0){2}}
\put(145,232.5){\vector(-1,0){2}}
\put(85,215){\vector(1,0){2}}
\put(145,215){\vector(1,0){2}}
\put(85,190){\vector(-1,0){2}}
\put(145,190){\vector(-1,0){2}}
\multiput(110,215)(3,5){5}{\line(0,1){5}}
\multiput(107,215)(3,5){6}{\line(1,0){3}}
\put(300,240){\line(1,0){90}}
\put(300,190){\line(1,0){90}}
\put(292,238){$c$}
\put(292,186){$\bar{s}$}
\put(273,210.5){$D^{+}_s$}
\put(393,238){$d$}
\put(393,186){$\bar{s}$}
\put(393,223){$\bar{s}$}
\put(393,200){$u$}
\put(410,193){$X^{+}_s$}
\put(410,230){$K^{(*)0}\Rightarrow K_{\rm S}$}
\put(390,215){\oval(70,25)[l]}
\put(315,240){\vector(1,0){2}}
\put(315,190){\vector(-1,0){2}}
\put(375,240){\vector(1,0){2}}
\put(375,190){\vector(-1,0){2}}
\put(375,227.5){\vector(-1,0){2}}
\put(375,202.5){\vector(1,0){2}}
\multiput(340,240)(3,-5){5}{\line(0,-1){5}}
\multiput(337,240)(3,-5){6}{\line(1,0){3}}
\end{picture}
\vspace{-6.2cm}
\caption{Doubly Cabibbo-suppressed channels for
$D^{+}\rightarrow K_{\rm S} X^+$, $(K_{\rm S} \pi^0)^{~}_{K^*} X^+$
and $D^+_s \rightarrow K_{\rm S} X^+_s$, $(K_{\rm S} \pi^0)^{~}_{K^*}
X^+_s$ decays
in the standard model.
Here $X = \pi$, $\rho$ or $a_1$,
$X_s = K$ or $K^*$, and $K_{\rm S}$ can
be replaced by $K_{\rm L}$.}
\end{figure}
 
\newpage

\begin{figure}
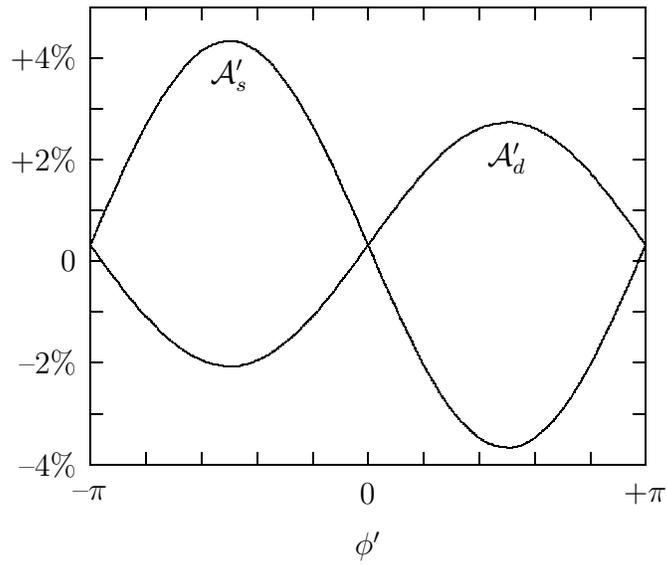

\setlength{\unitlength}{0.240900pt}
\ifx\plotpoint\undefined\newsavebox{\plotpoint}\fi
\sbox{\plotpoint}{\rule[-0.175pt]{0.350pt}{0.350pt}}%

\caption{Illustrative plot for $CP$ asymmetries ${\cal A}'_d$
and ${\cal A}'_s$ changing with the weak phase $\phi'$, where
$r' =0.04$, $R'_d \sin\delta'_d = 0.3$ and $R'_s \sin\delta'_s
=-0.5$ have typically been taken.}
\end{figure}
 
\end{document}